 \input pictex.tex   

\immediate\write10{Package DCpic 2002/05/16 v4.0}

\catcode`!=11 

\newcount\aux%
\newcount\auxa%
\newcount\auxb%
\newcount\m%
\newcount\n%
\newcount\x%
\newcount\y%
\newcount\xl%
\newcount\yl%
\newcount\d%
\newcount\dnm%
\newcount\xa%
\newcount\xb%
\newcount\xmed%
\newcount\xc%
\newcount\xd%
\newcount\ya%
\newcount\yb%
\newcount\ymed%
\newcount\yc%
\newcount\yd
\newcount\expansao%
\newcount\tipografo
\newcount\distanciaobjmor
\newcount\tipoarco
\newif\ifpara%
\newbox\caixa%
\newbox\caixaaux%
\newif\ifnvazia%
\newif\ifvazia%
\newif\ifcompara%
\newif\ifdiferentes%
\newcount\xaux%
\newcount\yaux%
\newcount\guardaauxa%
\newcount\alt%
\newcount\larg%
\newcount\prof%
\newcount\auxqx
\newcount\auxqy
\newif\ifajusta%
\newif\ifajustadist
\def\objPartida{}%
\def\objChegada{}%
\def\objNulo{}%


\def\!vazia{:}

\def\!pilhanvazia#1{\let\arg=#1%
\if:\arg\ \nvaziafalse\vaziatrue \else \nvaziatrue\vaziafalse\fi}

\def\!coloca#1#2{\edef\pilha{#1.#2}}

\def\!guarda(#1)(#2,#3)(#4,#5,#6){\def\id{#1}%
\xaux=#2%
\yaux=#3%
\alt=#4%
\larg=#5%
\prof=#6%
}

\def\!topaux#1.#2:{\!guarda#1}
\def\!topo#1{\expandafter\!topaux#1}

\def\!popaux#1.#2:{\def\pilha{#2:}}
\def\!retira#1{\expandafter\!popaux#1}

\def\!comparaaux#1#2{\let\argA=#1\let\argB=#2%
\ifx\argA\argB\comparatrue\diferentesfalse\else\comparafalse\diferentestrue\fi}

\def\!compara#1#2{\!comparaaux{#1}{#2}}

\def\!absoluto#1#2{\n=#1%
  \ifnum \n > 0
    #2=\n
  \else
    \multiply \n by -1
    #2=\n
  \fi}


\def\dasharrow{1}
\def\solidline{2}


\def\atleft{1}
\def\pup{0}

\def\commdiag{0}


\def\!ajusta#1#2#3#4#5#6{\aux=#5%
  \let\auxobj=#6%
  \ifcase \tipografo    
    \ifnum\number\aux=10 
      \ajustadisttrue 
    \else
      \ajustadistfalse  
    \fi
  \else  
   \ajustadistfalse
  \fi
  \ifajustadist
   \let\pilhaaux=\pilha%
   \loop%
     \!topo{\pilha}%
     \!retira{\pilha}%
     \!compara{\id}{\auxobj}%
     \ifcompara\nvaziafalse \else\!pilhanvazia\pilha \fi%
     \ifnvazia%
   \repeat%
   \let\pilha=\pilhaaux%
   \ifvazia%
    \ifdiferentes%
     \larg=1310720
     \prof=655360%
     \alt=655360%
    \fi%
   \fi%
   \divide\larg by 131072
   \divide\prof by 65536
   \divide\alt by 65536
   \ifnum\number\y=\number\yl
    \advance\larg by 3
    \ifnum\number\larg>\aux
     #5=\larg
    \fi
   \else
    \ifnum\number\x=\number\xl
     \ifnum\number\yl>\number\y
      \ifnum\number\alt>\aux
       #5=\alt
      \fi
     \else
      \advance\prof by 5
      \ifnum\number\prof>\aux
       #5=\prof
      \fi
     \fi
    \else
     \auxqx=\x
     \advance\auxqx by -\xl
     \!absoluto{\auxqx}{\auxqx}%
     \auxqy=\y
     \advance\auxqy by -\yl
     \!absoluto{\auxqy}{\auxqy}%
     \ifnum\auxqx>\auxqy
      \ifnum\larg<10
       \larg=10
      \fi
      \advance\larg by 3
      #5=\larg
     \else
      \ifnum\yl>\y
       \ifnum\larg<10
        \larg=10
       \fi
      \advance\alt by 6
       #5=\alt
      \else
      \advance\prof by 11
       #5=\prof
      \fi
     \fi
    \fi
   \fi
\fi} 

\def\!raiz#1#2{\n=#1%
  \m=1%
  \loop
    \aux=\m%
    \advance \aux by 1%
    \multiply \aux by \aux%
    \ifnum \aux < \n%
      \advance \m by 1%
      \paratrue%
    \else\ifnum \aux=\n%
      \advance \m by 1%
      \paratrue%
       \else\parafalse%
       \fi
    \fi
  \ifpara%
  \repeat
#2=\m}

\def\!ucoord#1#2#3#4#5#6#7{\aux=#2%
  \advance \aux by -#1%
  \multiply \aux by #4%
  \divide \aux by #5%
  \ifnum #7 = -1 \multiply \aux by -1 \fi%
  \advance \aux by #3%
#6=\aux}

\def\!quadrado#1#2#3{\aux=#1%
  \advance \aux by -#2%
  \multiply \aux by \aux%
#3=\aux}

\def\!distnomemor#1#2#3#4#5#6{\setbox0=\hbox{#5}%
  \aux=#1
  \advance \aux by -#3
  \ifnum \aux=0
     \aux=\wd0 \divide \aux by 131072
     \advance \aux by 3
     #6=\aux
  \else
     \aux=#2
     \advance \aux by -#4
     \ifnum \aux=0
        \aux=\ht0 \advance \aux by \dp0 \divide \aux by 131072
        \advance \aux by 3
        #6=\aux%
     \else
     #6=3
     \fi
   \fi
}

\def\begindc#1{\!ifnextchar[{\!begindc{#1}}{\!begindc{#1}[30]}}
\def\!begindc#1[#2]{\beginpicture 
  \let\pilha=\!vazia
  \setcoordinatesystem units <1pt,1pt>
  \expansao=#2
  \ifcase #1
    \distanciaobjmor=10
    \tipoarco=0         
    \tipografo=0        
  \or
    \distanciaobjmor=2
    \tipoarco=0         
    \tipografo=1        
  \or
    \distanciaobjmor=1
    \tipoarco=2         
    \tipografo=2        
  \or
    \distanciaobjmor=8
    \tipoarco=0         
    \tipografo=3        
  \or
    \distanciaobjmor=8
    \tipoarco=2         
    \tipografo=4        
  \fi}

\def\enddc{\endpicture}

\def\mor{%
  \!ifnextchar({\!morxy}{\!morObjA}}
\def\!morxy(#1,#2){%
  \!ifnextchar({\!morxyl{#1}{#2}}{\!morObjB{#1}{#2}}}
\def\!morxyl#1#2(#3,#4){%
  \!ifnextchar[{\!mora{#1}{#2}{#3}{#4}}{\!mora{#1}{#2}{#3}{#4}[\number\distanciaobjmor,\number\distanciaobjmor]}}%
\def\!morObjA#1{%
 \let\pilhaaux=\pilha%
 \def\objPartida{#1}%
 \loop%
    \!topo\pilha%
    \!retira\pilha%
    \!compara{\id}{\objPartida}%
    \ifcompara \nvaziafalse \else \!pilhanvazia\pilha \fi%
   \ifnvazia%
 \repeat%
 \ifvazia%
  \ifdiferentes%
   Error: Incorrect label specification%
   \xaux=1%
   \yaux=1%
  \fi%
 \fi%
 \let\pilha=\pilhaaux%
 \!ifnextchar({\!morxyl{\number\xaux}{\number\yaux}}{\!morObjB{\number\xaux}{\number\yaux}}}
\def\!morObjB#1#2#3{%
  \x=#1
  \y=#2
 \def\objChegada{#3}%
 \let\pilhaaux=\pilha%
 \loop
    \!topo\pilha %
    \!retira\pilha%
    \!compara{\id}{\objChegada}%
    \ifcompara \nvaziafalse \else \!pilhanvazia\pilha \fi
   \ifnvazia
 \repeat
 \ifvazia
  \ifdiferentes%
   Error: Incorrect label specification
   \xaux=\x%
   \advance\xaux by \x%
   \yaux=\y%
   \advance\yaux by \y%
  \fi
 \fi
 \let\pilha=\pilhaaux
 \!ifnextchar[{\!mora{\number\x}{\number\y}{\number\xaux}{\number\yaux}}{\!mora{\number\x}{\number\y}{\number\xaux}{\number\yaux}[\number\distanciaobjmor,\number\distanciaobjmor]}}
\def\!mora#1#2#3#4[#5,#6]#7{%
  \!ifnextchar[{\!morb{#1}{#2}{#3}{#4}{#5}{#6}{#7}}{\!morb{#1}{#2}{#3}{#4}{#5}{#6}{#7}[1,\number\tipoarco] }}
\def\!morb#1#2#3#4#5#6#7[#8,#9]{\x=#1%
  \y=#2%
  \xl=#3%
  \yl=#4%
  \multiply \x by \expansao%
  \multiply \y by \expansao%
  \multiply \xl by \expansao%
  \multiply \yl by \expansao%
  \!quadrado{\number\x}{\number\xl}{\auxa}%
  \!quadrado{\number\y}{\number\yl}{\auxb}%
  \d=\auxa%
  \advance \d by \auxb%
  \!raiz{\d}{\d}%
  \auxa=#5
  \!compara{\objNulo}{\objPartida}%
  \ifdiferentes
   \!ajusta{\x}{\xl}{\y}{\yl}{\auxa}{\objPartida}%
   \ajustatrue
   \def\objPartida{}
  \fi
  \guardaauxa=\auxa
  \!ucoord{\number\x}{\number\xl}{\number\x}{\auxa}{\number\d}{\xa}{1}%
  \!ucoord{\number\y}{\number\yl}{\number\y}{\auxa}{\number\d}{\ya}{1}%
  \auxa=\d%
  \auxb=#6
  \!compara{\objNulo}{\objChegada}%
  \ifdiferentes
   \!ajusta{\x}{\xl}{\y}{\yl}{\auxb}{\objChegada}%
   \def\objChegada{}
  \fi
  \advance \auxa by -\auxb%
  \!ucoord{\number\x}{\number\xl}{\number\x}{\number\auxa}{\number\d}{\xb}{1}%
  \!ucoord{\number\y}{\number\yl}{\number\y}{\number\auxa}{\number\d}{\yb}{1}%
  \xmed=\xa%
  \advance \xmed by \xb%
  \divide \xmed by 2
  \ymed=\ya%
  \advance \ymed by \yb%
  \divide \ymed by 2
  \!distnomemor{\number\x}{\number\y}{\number\xl}{\number\yl}{#7}{\dnm}%
  \!ucoord{\number\y}{\number\yl}{\number\xmed}{\number\dnm}{\number\d}{\xc}{-#8}%
  \!ucoord{\number\x}{\number\xl}{\number\ymed}{\number\dnm}{\number\d}{\yc}{#8}%
\ifcase #9  
  \arrow <4pt> [.2,1.1] from {\xa} {\ya} to {\xb} {\yb}
\or  
  \setdashes
  \arrow <4pt> [.2,1.1] from {\xa} {\ya} to {\xb} {\yb}
  \setsolid
\or  
  \setlinear
  \plot {\xa} {\ya}  {\xb} {\yb} /
\or  
  \auxa=\guardaauxa
  \advance \auxa by 3%
 \!ucoord{\number\x}{\number\xl}{\number\x}{\number\auxa}{\number\d}{\xa}{1}%
 \!ucoord{\number\y}{\number\yl}{\number\y}{\number\auxa}{\number\d}{\ya}{1}%
 \!ucoord{\number\y}{\number\yl}{\number\xa}{3}{\number\d}{\xd}{-1}%
 \!ucoord{\number\x}{\number\xl}{\number\ya}{3}{\number\d}{\yd}{1}%
  \arrow <4pt> [.2,1.1] from {\xa} {\ya} to {\xb} {\yb}
  \circulararc -180 degrees from {\xa} {\ya} center at {\xd} {\yd}
\or  
  \auxa=3
 \!ucoord{\number\y}{\number\yl}{\number\xa}{\number\auxa}{\number\d}{\xmed}{-1}%
 \!ucoord{\number\x}{\number\xl}{\number\ya}{\number\auxa}{\number\d}{\ymed}{1}%
 \!ucoord{\number\y}{\number\yl}{\number\xa}{\number\auxa}{\number\d}{\xd}{1}%
 \!ucoord{\number\x}{\number\xl}{\number\ya}{\number\auxa}{\number\d}{\yd}{-1}%
  \arrow <4pt> [.2,1.1] from {\xa} {\ya} to {\xb} {\yb}
  \setlinear
  \plot {\xmed} {\ymed}  {\xd} {\yd} /
\fi
\auxa=\xl
\advance \auxa by -\x%
\ifnum \auxa=0 
  \put {#7} at {\xc} {\yc}
\else
  \auxb=\yl
  \advance \auxb by -\y%
  \ifnum \auxb=0 \put {#7} at {\xc} {\yc}
  \else 
    \ifnum \auxa > 0 
      \ifnum \auxb > 0
        \ifnum #8=1
          \put {#7} [rb] at {\xc} {\yc}
        \else 
          \put {#7} [lt] at {\xc} {\yc}
        \fi
      \else
        \ifnum #8=1
          \put {#7} [lb] at {\xc} {\yc}
        \else 
          \put {#7} [rt] at {\xc} {\yc}
        \fi
      \fi
    \else
      \ifnum \auxb > 0 
        \ifnum #8=1
          \put {#7} [rt] at {\xc} {\yc}
        \else 
          \put {#7} [lb] at {\xc} {\yc}
        \fi
      \else
        \ifnum #8=1
          \put {#7} [lt] at {\xc} {\yc}
        \else 
          \put {#7} [rb] at {\xc} {\yc}
        \fi
      \fi
    \fi
  \fi
\fi
}

\def\modifplot(#1{\!modifqcurve #1}
\def\!modifqcurve(#1,#2){\x=#1%
  \y=#2%
  \multiply \x by \expansao%
  \multiply \y by \expansao%
  \!start (\x,\y)
  \!modifQjoin}
\def\!modifQjoin(#1,#2)(#3,#4){\x=#1%
  \y=#2%
  \xl=#3%
  \yl=#4%
  \multiply \x by \expansao%
  \multiply \y by \expansao%
  \multiply \xl by \expansao%
  \multiply \yl by \expansao%
  \!qjoin (\x,\y) (\xl,\yl)             
  \!ifnextchar){\!fim}{\!modifQjoin}}
\def\!fim){\ignorespaces}

\def\setaxy(#1{\!pontosxy #1}
\def\!pontosxy(#1,#2){%
  \!maispontosxy}
\def\!maispontosxy(#1,#2)(#3,#4){%
  \!ifnextchar){\!fimxy#3,#4}{\!maispontosxy}}
\def\!fimxy#1,#2){\x=#1%
  \y=#2
  \multiply \x by \expansao
  \multiply \y by \expansao
  \xl=\x%
  \yl=\y%
  \aux=1%
  \multiply \aux by \auxa%
  \advance\xl by \aux%
  \aux=1%
  \multiply \aux by \auxb%
  \advance\yl by \aux%
  \arrow <4pt> [.2,1.1] from {\x} {\y} to {\xl} {\yl}}

\def\cmor#1 #2(#3,#4)#5{%
  \!ifnextchar[{\!cmora{#1}{#2}{#3}{#4}{#5}}{\!cmora{#1}{#2}{#3}{#4}{#5}[0] }}
\def\!cmora#1#2#3#4#5[#6]{%
  \ifcase #2
      \auxa=0
      \auxb=1
    \or
      \auxa=0
      \auxb=-1
    \or
      \auxa=1
      \auxb=0
    \or
      \auxa=-1
      \auxb=0
    \fi  
  \ifcase #6  
    \modifplot#1
    \setaxy#1
  \or  
    \setdashes
    \modifplot#1
    \setaxy#1
    \setsolid
  \or  
    \modifplot#1
  \fi  
  \x=#3%
  \y=#4%
  \multiply \x by \expansao%
  \multiply \y by \expansao%
  \put {#5} at {\x} {\y}}

\def\obj(#1,#2){%
  \!ifnextchar[{\!obja{#1}{#2}}{\!obja{#1}{#2}[Nulo]}}
\def\!obja#1#2[#3]#4{%
  \!ifnextchar[{\!objb{#1}{#2}{#3}{#4}}{\!objb{#1}{#2}{#3}{#4}[1]}}
\def\!objb#1#2#3#4[#5]{%
  \x=#1%
  \y=#2%
  \def\!pinta{\normalsize$\bullet$}
  \def\!nulo{Nulo}%
  \def\!arg{#3}%
  \!compara{\!arg}{\!nulo}%
  \ifcompara\def\!arg{#4}\fi%
  \multiply \x by \expansao%
  \multiply \y by \expansao%
  \setbox\caixa=\hbox{#4}%
  \!coloca{(\!arg)(#1,#2)(\number\ht\caixa,\number\wd\caixa,\number\dp\caixa)}{\pilha}%
  \auxa=\wd\caixa \divide \auxa by 131072 
  \advance \auxa by 5
  \auxb=\ht\caixa
  \advance \auxb by \number\dp\caixa
  \divide \auxb by 131072 
  \advance \auxb by 5
  \ifcase \tipografo    
    \put{#4} at {\x} {\y}
  \or                   
    \ifcase #5 
      \put{#4} at {\x} {\y}
    \or        
      \put{\!pinta} at {\x} {\y}
      \advance \y by \number\auxb  
      \put{#4} at {\x} {\y}
    \or        
      \put{\!pinta} at {\x} {\y}
      \advance \auxa by -2  
      \advance \auxb by -2  
      \advance \x by \number\auxa  
      \advance \y by \number\auxb  
      \put{#4} at {\x} {\y}   
    \or        
      \put{\!pinta} at {\x} {\y}
      \advance \x by \number\auxa  
      \put{#4} at {\x} {\y}   
    \or        
      \put{\!pinta} at {\x} {\y}
      \advance \auxa by -2  
      \advance \auxb by -2  
      \advance \x by \number\auxa  
      \advance \y by -\number\auxb  
      \put{#4} at {\x} {\y}   
    \or        
      \put{\!pinta} at {\x} {\y}
      \advance \y by -\number\auxb  
      \put{#4} at {\x} {\y}   
    \or        
      \put{\!pinta} at {\x} {\y}
      \advance \auxa by -2  
      \advance \auxb by -2  
      \advance \x by -\number\auxa  
      \advance \y by -\number\auxb  
      \put{#4} at {\x} {\y}   
    \or        
      \put{\!pinta} at {\x} {\y}
      \advance \x by -\number\auxa  
      \put{#4} at {\x} {\y}   
    \or        
      \put{\!pinta} at {\x} {\y}
      \advance \auxa by -2  
      \advance \auxb by -2  
      \advance \x by -\number\auxa  
      \advance \y by \number\auxb  
      \put{#4} at {\x} {\y}   
    \fi
  \or                   
    \ifcase #5 
      \put{#4} at {\x} {\y}
    \or        
      \put{\!pinta} at {\x} {\y}
      \advance \y by \number\auxb  
      \put{#4} at {\x} {\y}
    \or        
      \put{\!pinta} at {\x} {\y}
      \advance \auxa by -2  
      \advance \auxb by -2  
      \advance \x by \number\auxa  
      \advance \y by \number\auxb  
      \put{#4} at {\x} {\y}   
    \or        
      \put{\!pinta} at {\x} {\y}
      \advance \x by \number\auxa  
      \put{#4} at {\x} {\y}   
    \or        
      \put{\!pinta} at {\x} {\y}
      \advance \auxa by -2  
      \advance \auxb by -2
      \advance \x by \number\auxa  
      \advance \y by -\number\auxb 
      \put{#4} at {\x} {\y}   
    \or        
      \put{\!pinta} at {\x} {\y}
      \advance \y by -\number\auxb 
      \put{#4} at {\x} {\y}   
    \or        
      \put{\!pinta} at {\x} {\y}
      \advance \auxa by -2  
      \advance \auxb by -2
      \advance \x by -\number\auxa 
      \advance \y by -\number\auxb 
      \put{#4} at {\x} {\y}   
    \or        
      \put{\!pinta} at {\x} {\y}
      \advance \x by -\number\auxa 
      \put{#4} at {\x} {\y}   
    \or        
      \put{\!pinta} at {\x} {\y}
      \advance \auxa by -2  
      \advance \auxb by -2
      \advance \x by -\number\auxa 
      \advance \y by \number\auxb  
      \put{#4} at {\x} {\y}   
    \fi
   \else 
     \ifnum\auxa<\auxb 
       \aux=\auxb
     \else
       \aux=\auxa
     \fi
     \ifdim\wd\caixa<1em
       \dimen99 = 1em
       \aux=\dimen99 \divide \aux by 131072 
       \advance \aux by 5
     \fi
     \advance\aux by -2 
     \multiply\aux by 2 %
     \ifnum\aux<30
       \put{\circle{\aux}} [Bl] at {\x} {\y}
     \else
       \multiply\auxa by 2
       \multiply\auxb by 2
       \put{\oval(\auxa,\auxb)} [Bl] at {\x} {\y}
     \fi
     \put{#4} at {\x} {\y}
   \fi   
}

\catcode`!=12 

  \input miniltx
  \def\Gin@driver{pdftex.def}
  \input color.sty
  \input graphicx.sty
  \resetatcatcode
%
%

%
%
%
%

\def\Serif{cmr}
\def\SerifBold{cmbx}
\def\SerifItalics{cmti}
\def\SerifSlanted{cmsl}
\def\SerifBoldItalics{cmbxti}
\def\SansSerif{cmss}
\def\SansSerifBold{cmssbx}
\def\SansSerifItalics{cmssi}
\def\SansSerifSlanted{cmssi}
\def\Math{cmmi}
\def\Symbols{cmsy}
\def\MathBold{cmmib}
\def\MoreSymbols{cmex}
\def\Typewriter{cmtt}
\def\Gothic{eufm}
\def\Double{msbm}

= 			\Serif10 			at 5pt
= 		\SerifBold10 		at 5pt
= 	\SerifItalics10 	at 5pt
=		\SerifSlanted10 	at 5pt
=	\SerifBoldItalics10	at 5pt
= 		\SansSerif10 		at 5pt
=	\SansSerifBold10	at 5pt
=	\SansSerifItalics10	at 5pt
=	\SansSerifSlanted10	at 5pt
=				\Math10				at 5pt
=			\MathBold10			at 5pt
=			\Symbols10			at 5pt
=		\MoreSymbols10		at 5pt
=		\Typewriter10		at 5pt
=			\Gothic10			at 5pt
=			\Double10			at 5pt

= 			\Serif10 			at 7pt
= 		\SerifBold10 		at 7pt
= 	\SerifItalics10 	at 7pt
=	\SerifSlanted10 	at 7pt
=\SerifBoldItalics10	at 7pt
= 		\SansSerif10 		at 7pt
= 	\SansSerifBold10 	at 7pt
=\SansSerifItalics10	at 7pt
=\SansSerifSlanted10	at 7pt
=			\Math10				at 7pt
=		\MathBold10			at 7pt
=			\Symbols10			at 7pt
=		\MoreSymbols10		at 7pt
=		\Typewriter10		at 7pt
=			\Gothic10			at 7pt
=			\Double10			at 7pt

= 			\Serif10 			at 8pt
= 		\SerifBold10 		at 8pt
= 	\SerifItalics10 	at 8pt
=	\SerifSlanted10 	at 8pt
=\SerifBoldItalics10	at 8pt
= 		\SansSerif10 		at 8pt
= 	\SansSerifBold10 	at 8pt
=\SansSerifItalics10 at 8pt
=\SansSerifSlanted10 at 8pt
=			\Math10				at 8pt
=		\MathBold10			at 8pt
=			\Symbols10			at 8pt
=		\MoreSymbols10		at 8pt
=		\Typewriter10		at 8pt
=			\Gothic10			at 8pt
=			\Double10			at 8pt

= 			\Serif10 			at 10pt
= 		\SerifBold10 		at 10pt
= 		\SerifItalics10 	at 10pt
=		\SerifSlanted10 	at 10pt
=	\SerifBoldItalics10	at 10pt
= 		\SansSerif10 		at 10pt
= 	\SansSerifBold10 	at 10pt
= 	\SansSerifItalics10 at 10pt
= 	\SansSerifSlanted10 at 10pt
=				\Math10				at 10pt
=			\MathBold10			at 10pt
=			\Symbols10			at 10pt
=		\MoreSymbols10		at 10pt
=		\Typewriter10		at 10pt
=			\Gothic10			at 10pt
=			\Double10			at 10pt

= 				\Serif10 			at 12pt
= 			\SerifBold10 		at 12pt
= 		\SerifItalics10 	at 12pt
=		\SerifSlanted10 	at 12pt
=	\SerifBoldItalics10	at 12pt
= 			\SansSerif10 		at 12pt
= 		\SansSerifBold10 	at 12pt
= 	\SansSerifItalics10 at 12pt
= 	\SansSerifSlanted10 at 12pt
=				\Math10				at 12pt
=			\MathBold10			at 12pt
=			\Symbols10			at 12pt
=		\MoreSymbols10		at 12pt
=			\Typewriter10		at 12pt
=				\Gothic10			at 12pt
=				\Double10			at 12pt

= 			\Serif10 			at 14pt
= 		\SerifBold10 		at 14pt
= 	\SerifItalics10 	at 14pt
=		\SerifSlanted10 	at 14pt
=	\SerifBoldItalics10	at 14pt
= 		\SansSerif10 		at 14pt
= 	\SansSerifBold10 	at 14pt
= \SansSerifSlanted10 at 14pt
= \SansSerifItalics10 at 14pt
=				\Math10				at 14pt
=			\MathBold10			at 14pt
=			\Symbols10			at 14pt
=		\MoreSymbols10		at 14pt
=		\Typewriter10		at 14pt
=			\Gothic10			at 14pt
=			\Double10			at 14pt

\def\NormalStyle{\parindent=5pt\parskip=3pt\normalbaselineskip=14pt%
\def\nt{\tenSerif}%
\def\rm{\fam0\tenSerif}%
\textfont0=\tenSerif\scriptfont0=\sevenSerif\scriptscriptfont0=\fiveSerif
\textfont1=\tenMath\scriptfont1=\sevenMath\scriptscriptfont1=\fiveMath
\textfont2=\tenSymbols\scriptfont2=\sevenSymbols\scriptscriptfont2=\fiveSymbols
\textfont3=\tenMoreSymbols\scriptfont3=\sevenMoreSymbols\scriptscriptfont3=\fiveMoreSymbols
\textfont\itfam=\tenSerifItalics\def\it{\fam\itfam\tenSerifItalics}%
\textfont\slfam=\tenSerifSlanted\def\sl{\fam\slfam\tenSerifSlanted}%
\textfont\ttfam=\tenTypewriter\def\tt{\fam\ttfam\tenTypewriter}%
\textfont\bffam=\tenSerifBold%
\def\bf{\fam\bffam\tenSerifBold}\scriptfont\bffam=\sevenSerifBold\scriptscriptfont\bffam=\fiveSerifBold%
\def\cal{\tenSymbols}%
\def\greekbold{\tenMathBold}%
\def\gothic{\tenGothic}%
\def\Bbb{\tenDouble}%
\def\LieFont{\tenSerifItalics}%
\nt\normalbaselines\baselineskip=14pt%
}

\def\TitleStyle{\parindent=0pt\parskip=0pt\normalbaselineskip=15pt%
\def\nt{\fourteenSansSerifBold}%
\def\rm{\fam0\fourteenSansSerifBold}%
\textfont0=\fourteenSansSerifBold\scriptfont0=\tenSansSerifBold\scriptscriptfont0=\eightSansSerifBold
\textfont1=\fourteenMath\scriptfont1=\tenMath\scriptscriptfont1=\eightMath
\textfont2=\fourteenSymbols\scriptfont2=\tenSymbols\scriptscriptfont2=\eightSymbols
\textfont3=\fourteenMoreSymbols\scriptfont3=\tenMoreSymbols\scriptscriptfont3=\eightMoreSymbols
\textfont\itfam=\fourteenSansSerifItalics\def\it{\fam\itfam\fourteenSansSerifItalics}%
\textfont\slfam=\fourteenSansSerifSlanted\def\sl{\fam\slfam\fourteenSerifSansSlanted}%
\textfont\ttfam=\fourteenTypewriter\def\tt{\fam\ttfam\fourteenTypewriter}%
\textfont\bffam=\fourteenSansSerif%
\def\bf{\fam\bffam\fourteenSansSerif}\scriptfont\bffam=\tenSansSerif\scriptscriptfont\bffam=\eightSansSerif%
\def\cal{\fourteenSymbols}%
\def\greekbold{\fourteenMathBold}%
\def\gothic{\fourteenGothic}%
\def\Bbb{\fourteenDouble}%
\def\LieFont{\fourteenSerifItalics}%
\nt\normalbaselines\baselineskip=15pt%
}

\def\PartStyle{\parindent=0pt\parskip=0pt\normalbaselineskip=15pt%
\def\nt{\fourteenSansSerifBold}%
\def\rm{\fam0\fourteenSansSerifBold}%
\textfont0=\fourteenSansSerifBold\scriptfont0=\tenSansSerifBold\scriptscriptfont0=\eightSansSerifBold
\textfont1=\fourteenMath\scriptfont1=\tenMath\scriptscriptfont1=\eightMath
\textfont2=\fourteenSymbols\scriptfont2=\tenSymbols\scriptscriptfont2=\eightSymbols
\textfont3=\fourteenMoreSymbols\scriptfont3=\tenMoreSymbols\scriptscriptfont3=\eightMoreSymbols
\textfont\itfam=\fourteenSansSerifItalics\def\it{\fam\itfam\fourteenSansSerifItalics}%
\textfont\slfam=\fourteenSansSerifSlanted\def\sl{\fam\slfam\fourteenSerifSansSlanted}%
\textfont\ttfam=\fourteenTypewriter\def\tt{\fam\ttfam\fourteenTypewriter}%
\textfont\bffam=\fourteenSansSerif%
\def\bf{\fam\bffam\fourteenSansSerif}\scriptfont\bffam=\tenSansSerif\scriptscriptfont\bffam=\eightSansSerif%
\def\cal{\fourteenSymbols}%
\def\greekbold{\fourteenMathBold}%
\def\gothic{\fourteenGothic}%
\def\Bbb{\fourteenDouble}%
\def\LieFont{\fourteenSerifItalics}%
\nt\normalbaselines\baselineskip=15pt%
}

\def\ChapterStyle{\parindent=0pt\parskip=0pt\normalbaselineskip=15pt%
\def\nt{\fourteenSansSerifBold}%
\def\rm{\fam0\fourteenSansSerifBold}%
\textfont0=\fourteenSansSerifBold\scriptfont0=\tenSansSerifBold\scriptscriptfont0=\eightSansSerifBold
\textfont1=\fourteenMath\scriptfont1=\tenMath\scriptscriptfont1=\eightMath
\textfont2=\fourteenSymbols\scriptfont2=\tenSymbols\scriptscriptfont2=\eightSymbols
\textfont3=\fourteenMoreSymbols\scriptfont3=\tenMoreSymbols\scriptscriptfont3=\eightMoreSymbols
\textfont\itfam=\fourteenSansSerifItalics\def\it{\fam\itfam\fourteenSansSerifItalics}%
\textfont\slfam=\fourteenSansSerifSlanted\def\sl{\fam\slfam\fourteenSerifSansSlanted}%
\textfont\ttfam=\fourteenTypewriter\def\tt{\fam\ttfam\fourteenTypewriter}%
\textfont\bffam=\fourteenSansSerif%
\def\bf{\fam\bffam\fourteenSansSerif}\scriptfont\bffam=\tenSansSerif\scriptscriptfont\bffam=\eightSansSerif%
\def\cal{\fourteenSymbols}%
\def\greekbold{\fourteenMathBold}%
\def\gothic{\fourteenGothic}%
\def\Bbb{\fourteenDouble}%
\def\LieFont{\fourteenSerifItalics}%
\nt\normalbaselines\baselineskip=15pt%
}

\def\SectionStyle{\parindent=0pt\parskip=0pt\normalbaselineskip=13pt%
\def\nt{\twelveSansSerifBold}%
\def\rm{\fam0\twelveSansSerifBold}%
\textfont0=\twelveSansSerifBold\scriptfont0=\eightSansSerifBold\scriptscriptfont0=\eightSansSerifBold
\textfont1=\twelveMath\scriptfont1=\eightMath\scriptscriptfont1=\eightMath
\textfont2=\twelveSymbols\scriptfont2=\eightSymbols\scriptscriptfont2=\eightSymbols
\textfont3=\twelveMoreSymbols\scriptfont3=\eightMoreSymbols\scriptscriptfont3=\eightMoreSymbols
\textfont\itfam=\twelveSansSerifItalics\def\it{\fam\itfam\twelveSansSerifItalics}%
\textfont\slfam=\twelveSansSerifSlanted\def\sl{\fam\slfam\twelveSerifSansSlanted}%
\textfont\ttfam=\twelveTypewriter\def\tt{\fam\ttfam\twelveTypewriter}%
\textfont\bffam=\twelveSansSerif%
\def\bf{\fam\bffam\twelveSansSerif}\scriptfont\bffam=\eightSansSerif\scriptscriptfont\bffam=\eightSansSerif%
\def\cal{\twelveSymbols}%
\def\bg{\twelveMathBold}%
\def\gothic{\twelveGothic}%
\def\Bbb{\twelveDouble}%
\def\LieFont{\twelveSerifItalics}%
\nt\normalbaselines\baselineskip=13pt%
}

\def\SubSectionStyle{\parindent=0pt\parskip=0pt\normalbaselineskip=13pt%
\def\nt{\twelveSansSerifItalics}%
\def\rm{\fam0\twelveSansSerifItalics}%
\textfont0=\twelveSansSerifItalics\scriptfont0=\eightSansSerifItalics\scriptscriptfont0=\eightSansSerifItalics%
\textfont1=\twelveMath\scriptfont1=\eightMath\scriptscriptfont1=\eightMath%
\textfont2=\twelveSymbols\scriptfont2=\eightSymbols\scriptscriptfont2=\eightSymbols%
\textfont3=\twelveMoreSymbols\scriptfont3=\eightMoreSymbols\scriptscriptfont3=\eightMoreSymbols%
\textfont\itfam=\twelveSansSerif\def\it{\fam\itfam\twelveSansSerif}%
\textfont\slfam=\twelveSansSerifSlanted\def\sl{\fam\slfam\twelveSerifSansSlanted}%
\textfont\ttfam=\twelveTypewriter\def\tt{\fam\ttfam\twelveTypewriter}%
\textfont\bffam=\twelveSansSerifBold%
\def\bf{\fam\bffam\twelveSansSerifBold}\scriptfont\bffam=\eightSansSerifBold\scriptscriptfont\bffam=\eightSansSerifBold%
\def\cal{\twelveSymbols}%
\def\greekbold{\twelveMathBold}%
\def\gothic{\twelveGothic}%
\def\Bbb{\twelveDouble}%
\def\LieFont{\twelveSerifItalics}%
\nt\normalbaselines\baselineskip=13pt%
}

\def\AuthorStyle{\parindent=0pt\parskip=0pt\normalbaselineskip=14pt%
\def\nt{\tenSerif}%
\def\rm{\fam0\tenSerif}%
\textfont0=\tenSerif\scriptfont0=\sevenSerif\scriptscriptfont0=\fiveSerif
\textfont1=\tenMath\scriptfont1=\sevenMath\scriptscriptfont1=\fiveMath
\textfont2=\tenSymbols\scriptfont2=\sevenSymbols\scriptscriptfont2=\fiveSymbols
\textfont3=\tenMoreSymbols\scriptfont3=\sevenMoreSymbols\scriptscriptfont3=\fiveMoreSymbols
\textfont\itfam=\tenSerifItalics\def\it{\fam\itfam\tenSerifItalics}%
\textfont\slfam=\tenSerifSlanted\def\sl{\fam\slfam\tenSerifSlanted}%
\textfont\ttfam=\tenTypewriter\def\tt{\fam\ttfam\tenTypewriter}%
\textfont\bffam=\tenSerifBold%
\def\bf{\fam\bffam\tenSerifBold}\scriptfont\bffam=\sevenSerifBold\scriptscriptfont\bffam=\fiveSerifBold%
\def\cal{\tenSymbols}%
\def\greekbold{\tenMathBold}%
\def\gothic{\tenGothic}%
\def\Bbb{\tenDouble}%
\def\LieFont{\tenSerifItalics}%
\nt\normalbaselines\baselineskip=14pt%
}

\def\AddressStyle{\parindent=0pt\parskip=0pt\normalbaselineskip=14pt%
\def\nt{\eightSerif}%
\def\rm{\fam0\eightSerif}%
\textfont0=\eightSerif\scriptfont0=\sevenSerif\scriptscriptfont0=\fiveSerif
\textfont1=\eightMath\scriptfont1=\sevenMath\scriptscriptfont1=\fiveMath
\textfont2=\eightSymbols\scriptfont2=\sevenSymbols\scriptscriptfont2=\fiveSymbols
\textfont3=\eightMoreSymbols\scriptfont3=\sevenMoreSymbols\scriptscriptfont3=\fiveMoreSymbols
\textfont\itfam=\eightSerifItalics\def\it{\fam\itfam\eightSerifItalics}%
\textfont\slfam=\eightSerifSlanted\def\sl{\fam\slfam\eightSerifSlanted}%
\textfont\ttfam=\eightTypewriter\def\tt{\fam\ttfam\eightTypewriter}%
\textfont\bffam=\eightSerifBold%
\def\bf{\fam\bffam\eightSerifBold}\scriptfont\bffam=\sevenSerifBold\scriptscriptfont\bffam=\fiveSerifBold%
\def\cal{\eightSymbols}%
\def\greekbold{\eightMathBold}%
\def\gothic{\eightGothic}%
\def\Bbb{\eightDouble}%
\def\LieFont{\eightSerifItalics}%
\nt\normalbaselines\baselineskip=14pt%
}

\def\AbstractStyle{\parindent=0pt\parskip=0pt\normalbaselineskip=12pt%
\def\nt{\eightSerif}%
\def\rm{\fam0\eightSerif}%
\textfont0=\eightSerif\scriptfont0=\sevenSerif\scriptscriptfont0=\fiveSerif
\textfont1=\eightMath\scriptfont1=\sevenMath\scriptscriptfont1=\fiveMath
\textfont2=\eightSymbols\scriptfont2=\sevenSymbols\scriptscriptfont2=\fiveSymbols
\textfont3=\eightMoreSymbols\scriptfont3=\sevenMoreSymbols\scriptscriptfont3=\fiveMoreSymbols
\textfont\itfam=\eightSerifItalics\def\it{\fam\itfam\eightSerifItalics}%
\textfont\slfam=\eightSerifSlanted\def\sl{\fam\slfam\eightSerifSlanted}%
\textfont\ttfam=\eightTypewriter\def\tt{\fam\ttfam\eightTypewriter}%
\textfont\bffam=\eightSerifBold%
\def\bf{\fam\bffam\eightSerifBold}\scriptfont\bffam=\sevenSerifBold\scriptscriptfont\bffam=\fiveSerifBold%
\def\cal{\eightSymbols}%
\def\greekbold{\eightMathBold}%
\def\gothic{\eightGothic}%
\def\Bbb{\eightDouble}%
\def\LieFont{\eightSerifItalics}%
\nt\normalbaselines\baselineskip=12pt%
}

\def\RefsStyle{\parindent=0pt\parskip=0pt%
\def\nt{\eightSerif}%
\def\rm{\fam0\eightSerif}%
\textfont0=\eightSerif\scriptfont0=\sevenSerif\scriptscriptfont0=\fiveSerif
\textfont1=\eightMath\scriptfont1=\sevenMath\scriptscriptfont1=\fiveMath
\textfont2=\eightSymbols\scriptfont2=\sevenSymbols\scriptscriptfont2=\fiveSymbols
\textfont3=\eightMoreSymbols\scriptfont3=\sevenMoreSymbols\scriptscriptfont3=\fiveMoreSymbols
\textfont\itfam=\eightSerifItalics\def\it{\fam\itfam\eightSerifItalics}%
\textfont\slfam=\eightSerifSlanted\def\sl{\fam\slfam\eightSerifSlanted}%
\textfont\ttfam=\eightTypewriter\def\tt{\fam\ttfam\eightTypewriter}%
\textfont\bffam=\eightSerifBold%
\def\bf{\fam\bffam\eightSerifBold}\scriptfont\bffam=\sevenSerifBold\scriptscriptfont\bffam=\fiveSerifBold%
\def\cal{\eightSymbols}%
\def\greekbold{\eightMathBold}%
\def\gothic{\eightGothic}%
\def\Bbb{\eightDouble}%
\def\LieFont{\eightSerifItalics}%
\nt\normalbaselines\baselineskip=10pt%
}



%
%


\def\ModeYes{yes}
\def\ModeNo{no}

\def\ModeUndef{undefined}


\def\nx{\noexpand}
\def\ni{\noindent}
\def\newpage{\vfill\eject}

\def\ss{\vskip 5pt}
\def\ms{\vskip 10pt}
\def\bs{\vskip 20pt}

 \def\,{\mskip\thinmuskip}
 \def\!{\mskip-\thinmuskip}
 \def\>{\mskip\medmuskip}
 \def\;{\mskip\thickmuskip}

%
% ... 
%

\def\refsModePost{post}
\def\refsModeAuto{auto}

\def\dbRefsSatusModeOk{ok}
\def\dbRefsSatusModeError{error}
\def\dbRefsSatusModeWarning{warning}


\newcount\BNUM
\BNUM=0

\def\refs{}

\def\SetModePost{\xdef\refsMode{\refsModePost}}			
\SetModePost

\def\dbRefsStatusOk{%
	\xdef\dbRefsStatus{\dbRefsSatusModeOk}%
	\xdef\dbRefsError{\ModeNo}%
	\xdef\dbRefsWarning{\ModeNo}%
	\xdef\dbRefsInfo{\ModeNo}%
}

\def\dbRefs{%
}

\def\dbRefsGet#1{%
	\xdef\found{N}\xdef\ikey{#1}\dbRefsStatusOk%
	\xdef\key{\ModeUndef}\xdef\tag{\ModeUndef}\xdef\tail{\ModeUndef}%
	\dbRefs%
}

\def\NextRefsTag{%
	\global\advance\BNUM by 1%
}
\def\ShowTag#1{{\bf [#1]}}

\def\dbRefsInsert#1#2{%
\dbRefsGet{#1}%
\if\found Y %
   \xdef\dbRefsStatus{\dbRefsSatusModeWarning}%
   \xdef\dbRefsWarning{record is already there}%
   \xdef\dbRefsInfo{record not inserted}%
\else%
   \toks2=\expandafter{\dbRefs}%
   \ifx\refsMode\refsModeAuto \NextRefsTag
    \xdef\dbRefs{%
   	\the\toks2 \nx\xdef\nx\dbx{#1}%
	\nx\ifx\nx\ikey %
		\nx\dbx\nx\xdef\nx\found{Y}%
		\nx\xdef\nx\key{#1}%
		\nx\xdef\nx\tag{\the\BNUM}%
		\nx\xdef\nx\tail{#2}%
	\nx\fi}%
	\global\xdef\refs{\refs \ss\ni[\the\BNUM]\ #2\par}
   \fi%
   \ifx\refsMode\refsModePost 
    \xdef\dbRefs{%
   	\the\toks2 \nx\xdef\nx\dbx{#1}%
	\nx\ifx\nx\ikey %
		\nx\dbx\nx\xdef\nx\found{Y}%
		\nx\xdef\nx\key{#1}%
		\nx\xdef\nx\tag{\ModeUndef}%
		\nx\xdef\nx\tail{#2}%
	\nx\fi}%
   \fi%
\fi%
}

\def\dbRefsEdit#1#2#3{\dbRefsGet{#1}%
\if\found N 
   \xdef\dbRefsStatus{\dbRefsSatusModeError}%
   \xdef\dbRefsError{record is not there}%
   \xdef\dbRefsInfo{record not edited}%
\else%
   \toks2=\expandafter{\dbRefs}%
   \xdef\dbRefs{\the\toks2%
   \nx\xdef\nx\dbx{#1}%
   \nx\ifx\nx\ikey\nx\dbx %
	\nx\xdef\nx\found{Y}%
	\nx\xdef\nx\key{#1}%
	\nx\xdef\nx\tag{#2}%
	\nx\xdef\nx\tail{#3}%
   \nx\fi}%
\fi%
}

\def\bib#1#2{\RefsStyle\dbRefsInsert{#1}{#2}%
	\ifx\dbRefsStatus\dbRefsSatusModeWarning %
		\message{^^J}%
		\message{WARNING: Reference [#1] is doubled.^^J}%
	\fi%
}

\def\ref#1{\dbRefsGet{#1}%
\ifx\found N %
  \message{^^J}%
  \message{ERROR: Reference [#1] unknown.^^J}%
  \ShowTag{??}%
\else%
	\ifx\tag\ModeUndef \NextRefsTag%
		\dbRefsEdit{#1}{\the\BNUM}{\tail}%
		\dbRefsGet{#1}%
		\global\xdef\refs{\refs \ss\ni [\tag]\ \tail\par}
	\fi
	\ShowTag{\tag}%
\fi%
}

\def\ShowBiblio{\bs\Ensure{\SectionEnsure}%
{\SectionStyle\ni References}%
{\RefsStyle\refs}%
}

\newcount\CHANGES
\CHANGES=0
\def\AuxFile{7}
\def\PreventDoubleOn{\xdef\PreventDoubleLabel{\ModeYes}}

\PreventDoubleOn

\def\StoreLabel#1#2{\xdef\itag{#2}
 \ifx\PreModeStatus\ModeNo %
   \message{^^J}%
   \errmessage{You can't use Check without starting with OpenPreMode (and finishing with ClosePreMode)^^J}%
 \else%
   \immediate\write\AuxFile{\nx\dbLabelPreInsert{#1}{\itag}}%
   \dbLabelGet{#1}%
   \ifx\itag\tag %
   \else%
	\global\advance\CHANGES by 1%
 	\xdef\itag{(?.??)}%
    \fi%
   \fi%
}

\def\PreModeStatus{\ModeNo}

\def\ClosePreMode{\immediate\closeout\AuxFile%
  \ifnum\CHANGES=0%
	\message{^^J}%
	\message{**********************************^^J}%
	\message{**  NO CHANGES TO THE AuxFile  **^^J}%
	\message{**********************************^^J}%
 \else%
	\message{^^J}%
	\message{**************************************************^^J}%
	\message{**  PLAEASE TYPESET IT AGAIN (\the\CHANGES)  **^^J}%
    \errmessage{**************************************************^^ J}%
  \fi%
  \edef\PreModeStatus{\ModeNo}%
}

\def\dbLabelSatusModeOk{ok}

\def\dbLabelSatusModeWarning{warning}

\def\dbLabelStatusOk{%
	\xdef\dbLabelStatus{\dbLabelSatusModeOk}%
	\xdef\dbLabelError{\ModeNo}%
	\xdef\dbLabelWarning{\ModeNo}%
	\xdef\dbLabelInfo{\ModeNo}%
}

\def\dbLabel{%
}

\def\dbLabelGet#1{%
	\xdef\found{N}\xdef\ikey{#1}\dbLabelStatusOk%
	\xdef\key{\ModeUndef}\xdef\tag{\ModeUndef}\xdef\pre{\ModeUndef}%
	\dbLabel%
}

\def\ShowLabel#1{%
 \dbLabelGet{#1}%
 \ifx\tag \ModeUndef %
 	\global\advance\CHANGES by 1%
 	(?.??)%
 \else%
 	\tag%
 \fi%
}

\def\dbLabelPreInsert#1#2{\dbLabelGet{#1}%
\if\found Y %
  \xdef\dbLabelStatus{\dbLabelSatusModeWarning}%
   \xdef\dbLabelWarning{Label is already there}%
   \xdef\dbLabelInfo{Label not inserted}%
   \message{^^J}%
   \errmessage{Double pre definition of label [#1]^^J}%
\else%
   \toks2=\expandafter{\dbLabel}%
    \xdef\dbLabel{%
   	\the\toks2 \nx\xdef\nx\dbx{#1}%
	\nx\ifx\nx\ikey %
		\nx\dbx\nx\xdef\nx\found{Y}%
		\nx\xdef\nx\key{#1}%
		\nx\xdef\nx\tag{#2}%
		\nx\xdef\nx\pre{\ModeYes}%
	\nx\fi}%
\fi%
}

\def\dbLabelInsert#1#2{\dbLabelGet{#1}%
\xdef\itag{#2}%
\dbLabelGet{#1}%
\if\found Y %
	\ifx\tag\itag %
	\else%
	   \ifx\PreventDoubleLabel\ModeYes %
		\message{^^J}%
		\errmessage{Double definition of label [#1]^^J}%
	   \else%
		\message{^^J}%
		\message{Double definition of label [#1]^^J}%
	   \fi%
	\fi%
   \xdef\dbLabelStatus{\dbLabelSatusModeWarning}%
   \xdef\dbLabelWarning{Label is already there}%
   \xdef\dbLabelInfo{Label not inserted}%
\else%
   \toks2=\expandafter{\dbLabel}%
    \xdef\dbLabel{%
   	\the\toks2 \nx\xdef\nx\dbx{#1}%
	\nx\ifx\nx\ikey %
		\nx\dbx\nx\xdef\nx\found{Y}%
		\nx\xdef\nx\key{#1}%
		\nx\xdef\nx\tag{#2}%
		\nx\xdef\nx\pre{\ModeNo}%
	\nx\fi}%
\fi%
}


\newcount\PART
\newcount\CHAPTER
\newcount\SECTION
\newcount\SUBSECTION
\newcount\FNUMBER

\PART=0
\CHAPTER=0
\SECTION=0
\SUBSECTION=0	
\FNUMBER=0

\def\LastPart{\ModeUndef}
\def\LastChapter{\ModeUndef}
\def\LastSection{\ModeUndef}
\def\LastSubSection{\ModeUndef}
\def\LastClaim{\ModeUndef}
\def\Last{\ModeUndef}

\newdimen\TOBOTTOM
\newdimen\LIMIT

\def\Ensure#1{\ \par\ \immediate\LIMIT=#1\immediate\TOBOTTOM=\the\pagegoal\advance\TOBOTTOM by -\pagetotal%
\ifdim\TOBOTTOM<\LIMIT\newpage \else%
\vskip-\parskip\vskip-\parskip\vskip-\baselineskip\fi}

\def\PartLabel{\the\PART}
\def\NewPart#1{\global\advance\PART by 1%
         \bs\ni{\PartStyle  Part \PartLabel:}
         \bs\ni{\PartStyle #1}\newpage%
         \CHAPTER=0\SECTION=0\SUBSECTION=0\FNUMBER=0%
         \gdef\Left{#1}%
         \global\edef\Last{\PartLabel}%
         \global\edef\LastPart{\PartLabel}%
         \global\edef\LastChapter{\ModeUndef}%
         \global\edef\LastSection{\ModeUndef}%
         \global\edef\LastSubSection{\ModeUndef}%
         \global\edef\LastClaim{\ModeUndef}}
\def\ChapterLabel{\the\CHAPTER}
\def\NewChapter#1{\global\advance\CHAPTER by 1%
         \bs\ni{\ChapterStyle  Chapter \ChapterLabel: #1}\ms%
         \SECTION=0\SUBSECTION=0\FNUMBER=0%
         \gdef\Left{#1}%
         \global\edef\Last{\ChapterLabel}%
         \global\edef\LastChapter{\ChapterLabel}%
         \global\edef\LastSection{\ModeUndef}%
         \global\edef\LastSubSection{\ModeUndef}%
         \global\edef\LastClaim{\ModeUndef}}
\def\SectionEnsure{3cm}
\def\NewSection#1{\Ensure{\SectionEnsure}\gdef\SectionLabel{\the\SECTION}\global\advance\SECTION by 1%
         \bs\ni{\SectionStyle  \SectionLabel.\ #1}\ss%
         \SUBSECTION=0\FNUMBER=0%
         \gdef\Left{#1}%
         \global\edef\Last{\SectionLabel}%
         \global\edef\LastSection{\SectionLabel}%
         \global\edef\LastSubSection{\ModeUndef}%
         \global\edef\LastClaim{\ModeUndef}}
\def\NewAppendix#1#2{\Ensure{\SectionEnsure}\gdef\SectionLabel{#1}\global\advance\SECTION by 1%
         \bs\ni{\SectionStyle  Appendix \SectionLabel.\ #2}\ss%
         \SUBSECTION=0\FNUMBER=0%
         \gdef\Left{#2}%
         \global\edef\Last{\SectionLabel}%
         \global\edef\LastSection{\SectionLabel}%
         \global\edef\LastSubSection{\ModeUndef}%
         \global\edef\LastClaim{\ModeUndef}}
\def\Acknowledgements{\Ensure{\SectionEnsure}\gdef\SectionLabel{}%
         \bs\ni{\SectionStyle  Acknowledgments}\ss%
         \SECTION=0\SUBSECTION=0\FNUMBER=0%
         \gdef\Left{}%
         \global\edef\Last{\ModeUndef}%
         \global\edef\LastSection{\ModeUndef}%
         \global\edef\LastSubSection{\ModeUndef}%
         \global\edef\LastClaim{\ModeUndef}}
\def\SubSectionEnsure{2cm}
\def\SubSectionLabel{\ifnum\SECTION>0 \the\SECTION.\fi\the\SUBSECTION}
\def\NewSubSection#1{\Ensure{\SubSectionEnsure}\global\advance\SUBSECTION by 1%
         \ms\ni{\SubSectionStyle #1}\ss%
         \global\edef\Last{\SubSectionLabel}%
         \global\edef\LastSubSection{\SubSectionLabel}}
\def\SetNumberingModeN{\def\ClaimLabel{(\the\FNUMBER)}}
\def\SetNumberingModeSN{\def\ClaimLabel{(\ifnum\SECTION>0 \SectionLabel.\fi%
      \the\FNUMBER)}}
\def\SetNumberingModeCSN{\def\ClaimLabel{(\ifnum\CHAPTER>0 \the\CHAPTER.\fi%
      \ifnum\SECTION>0 \SectionLabel.\fi%
      \the\FNUMBER)}}

\def\NewClaim{\global\advance\FNUMBER by 1%
    \ClaimLabel%
    \global\edef\LastClaim{\ClaimLabel}%
    \global\edef\Last{\ClaimLabel}}

\def\HideLabels{\xdef\ShowLabelsMode{\ModeNo}}
\HideLabels

\def\fn{\eqno{\NewClaim}} 
\def\fl#1{%
\ifx\ShowLabelsMode\ModeYes%
 \eqno{{\buildrel{\hbox{\AbstractStyle[#1]}}\over{\hfill\NewClaim}}}%
\else%
 \eqno{\NewClaim}%
\fi%
\dbLabelInsert{#1}{\ClaimLabel}}
\def\fprel#1{\global\advance\FNUMBER by 1\StoreLabel{#1}{\ClaimLabel}%
\ifx\ShowLabelsMode\ModeYes%
\eqno{{\buildrel{\hbox{\AbstractStyle[#1]}}\over{\hfill.\itag}}}%
\else%
 \eqno{\itag}%
\fi%
}

\def\cl#1{\global\advance\FNUMBER by 1\dbLabelInsert{#1}{\ClaimLabel}%
\ifx\ShowLabelsMode\ModeYes%
${\buildrel{\hbox{\AbstractStyle[#1]}}\over{\hfill\ClaimLabel}}$%
\else%
  $\ClaimLabel$%
\fi%
}
\def\cprel#1{\global\advance\FNUMBER by 1\StoreLabel{#1}{\ClaimLabel}%
\ifx\ShowLabelsMode\ModeYes%
${\buildrel{\hbox{\AbstractStyle[#1]}}\over{\hfill.\itag}}$%
\else%
  $\itag$%
\fi%
}


\parindent=7pt
\leftskip=2cm
\newcount\SideIndent
\newcount\SideIndentTemp
\SideIndent=0
\newdimen\SectionIndent
\SectionIndent=-8pt

\def\sidebar{\vrule height15pt width.2pt }
\def\endcorner{\hbox{\hbox{\vrule height6pt width.2pt}\vbox to6pt{\vfill\hbox
to4pt{\leaders\hrule height0.2pt\hfill}}}}
\def\begincorner{\hbox{\hbox{\vrule height6pt width.2pt}\vbox to6pt{\hbox
to4pt{\leaders\hrule height0.2pt\hfill}}}}
\def\endbegincorner{\hbox{\vbox to15pt{\endcorner\vskip-6pt\begincorner\vfill}}}
\def\SideShow{\SideIndentTemp=\SideIndent \ifnum \SideIndentTemp>0 
\loop\sidebar\hskip 2pt \advance\SideIndentTemp by-1\ifnum \SideIndentTemp>1 \repeat\fi}

\def\BeginSection{{\vbadness 100000 \par\ni\hskip\SectionIndent%
\SideShow\vbox to 15pt{\vfill\begincorner}}\global\advance\SideIndent by1\vskip-10pt}

\def\EndSection{{\vbadness 100000 \par\ni\global\advance\SideIndent by-1%
\hskip\SectionIndent\SideShow\vbox to15pt{\endcorner\vfill}\vskip-10pt}}

\def\EndBeginSection{{\vbadness 100000\par\ni%
\global\advance\SideIndent by-1\hskip\SectionIndent\SideShow
\vbox to15pt{\vfill\endbegincorner}}%
\global\advance\SideIndent by1\vskip-10pt}

\def\ShowBeginCorners#1{%
\SideIndentTemp =#1 \advance\SideIndentTemp by-1%
\ifnum \SideIndentTemp>0 %
\vskip-15truept\hbox{\kern 2truept\vbox{\hbox{\begincorner}%
\ShowBeginCorners{\SideIndentTemp}\vskip-3truept}}%
\fi%
}

\def\ShowEndCorners#1{%
\SideIndentTemp =#1 \advance\SideIndentTemp by-1%
\ifnum \SideIndentTemp>0 %
\vskip-15truept\hbox{\kern 2truept\vbox{\hbox{\endcorner}%
\ShowEndCorners{\SideIndentTemp}\vskip 2truept}}%
\fi%
}

\def\BeginSections#1{{\vbadness 100000 \par\ni\hskip\SectionIndent%
\SideShow\vbox to 15pt{\vfill\ShowBeginCorners{#1}}}\global\advance\SideIndent by#1\vskip-10pt}

\def\EndSections#1{{\vbadness 100000 \par\ni\global\advance\SideIndent by-#1%
\hskip\SectionIndent\SideShow\vbox to15pt{\vskip15pt\ShowEndCorners{#1}\vfill}\vskip-10pt}}

\def\EndBeginSections#1#2{{\vbadness 100000\par\ni%
\global\advance\SideIndent by-#1%
\hbox{\hskip\SectionIndent\SideShow\kern-2pt%
\vbox to15pt{\vskip15pt\ShowEndCorners{#1}\vskip4pt\ShowBeginCorners{#2}}}}%
\global\advance\SideIndent by#2\vskip-10pt}



\def\Note{\ms\leftskip 3cm\rightskip 1.5cm\AbstractStyle}
\def\endNote{\par\leftskip 2cm\rightskip 0cm\NormalStyle\ss}

\def\CollapseAllCNotes{\long\def\CNote##1{}}
\def\ExpandAllCNotes{\long\def\CNote##1{%
\BeginSection
	\Note%
 		##1%
	\endNote%
\EndSection%
}}
\ExpandAllCNotes


\def\frame#1{\vbox{\hrule\hbox{\vrule\vbox{\kern2pt\hbox{\kern2pt#1\kern2pt}\kern2pt}\vrule}\hrule\kern-4pt}}

\def\uline#1{\underline{#1}}
\def\uuline#1{\underline{\underline{#1}}}
\def\Box to #1#2#3{\frame{\vtop{\hbox to #1{\hfill #2 \hfill}\hbox to #1{\hfill #3 \hfill}}}}


%
%


\def\al{\alpha}
\def\be{\beta}
\def\de{\delta}
\def\ga{\gamma}

\def\ep{\epsilon}
\def\io{\iota}

\def\la{\lambda}

\def\om{\omega}
\def\si{\sigma}

\def\etsme #1#2{\hbox{diag}(\underbrace{1,\dots, 1}_{{#1} \rm \; times}, \underbrace{-1,\dots, -1}_{{#2} \rm \; times})}
\def\retsme #1#2{\hbox{diag}(\underbrace{-1,\dots, -1}_{{#1} \rm \; times}, \underbrace{1,\dots, 1}_{{#2} \rm \; times})}


 \def\calC{{\hbox{\cal C}}}


 \def\gotg{{\hbox{\gothic g}}}
 \def\goth{{\hbox{\gothic h}}}
 
 \def\spin{{\hbox{\gothic spin}}}

 \def\one{{\hbox{\Bbb I}}}
 
 \def\R{{\hbox{\Bbb R}}}

 \def\R{{\hbox{\Bbb R}}}


\def\ad{{\hbox{ad}}}

\def\Spin{{\hbox{Spin}}}
\def\SO{{\hbox{SO}}}
\def\SU{{\hbox{SU}}}

\def\id{{\hbox{\rm id}}}
\def\diag{{\hbox{\rm diag}}}

\def\ip{\hbox to4pt{\leaders\hrule height0.3pt\hfill}\vbox to8pt{\leaders\vrule width0.3pt\vfill}\kern 2pt}

\def\arr{\rightarrow}

%
%

\long\def\title#1{\centerline{\TitleStyle\ni#1}}
\long\def\moretitle#1{\centerline{\TitleStyle\ni#1}}
\long\def\author#1{\ms\centerline{\AuthorStyle by {\it #1}}}

\long\def\address#1{\ss\centerline{\AddressStyle #1}\par}
\long\def\moreaddress#1{\centerline{\AddressStyle #1}\par}
\def\abstract{\ms\leftskip 3cm\rightskip .5cm\AbstractStyle{\bf \ni Abstract:}\ }
\def\endabstract{\par\leftskip 2cm\rightskip 0cm\NormalStyle\ss}

\def\cases#1{\left\{\eqalign{#1}\right.}
\NormalStyle
\SetNumberingModeSN
\PreventDoubleOn

\SetNumberingModeSN

\def\so{{\gothic so}}

\def\frac[#1/#2]{\hbox{$#1\over#2$}}

\def\({\left(}
\def\){\right)}
\def\[{\left[}
\def\]{\right]}
\def\^#1{{}^{#1}_{\>\cdot}}
\def\_#1{{}_{#1}^{\>\cdot}}
\def\Label=#1{{\buildrel {\hbox{\fiveSerif \ShowLabel{#1}}}\over =}}
\def\<{\kern -1pt}

\bib{Book}{L.\ Fatibene, M.\ Francaviglia, 
{\it Natural and gauge natural formalism for classical field theories. A geometric perspective including spinors and gauge theories}, 
Kluwer Academic Publishers, Dordrecht, 2003}

\bib{R1}{L. Fatibene, M.Francaviglia, C.Rovelli, {\it On a Covariant Formulation of the Barberi-Immirzi Connection}
CQG 24 (2007) 3055-3066; gr-qc/0702134}

\bib{R2}{L. Fatibene, M.Francaviglia, C.Rovelli, {\it Spacetime Lagrangian Formulation of Barbero-Immirzi Gravity} 
CQG 24 (2007) 4207-4217; gr-qc/0706.1899}

\bib{Smirnov}{A.L. Smirnov, (private communication)}

\bib{Samuel}{J.\ Samuel, {\it Is Barbero's Hamiltonian Formulation a Gauge Theory of Lorentzian Gravity?}, Class.\ Quantum Grav.\ {\it 17}, 2000, 141-148}

\bib{Barbero}{F.\ Barbero, {\it Real Ashtekar variables for Lorentzian signature space-time},
Phys.\ Rev.\ {\it D51}, 5507, 1996}

\bib{Immirzi}{G.\ Immirzi, {\it Quantum Gravity and Regge Calculus},
Nucl.\ Phys.\ Proc.\ Suppl.\ {\bf 57}, 65-72}

\bib{RovelliBook}{C.\ Rovelli, {\it Quantum Gravity}, Cambridge University Press, Cambridge, 2004}

\bib{Thie2006}{T. Thiemann,
{\it Loop Quantum Gravity: An Inside View}, hep-th/0608210}

\bib{Gatto}{M.Ferraris, M.Francaviglia, L.Gatto, 
{\it Reducibility of $G$-invariant Linear Connections in Principal $G$-bundles}, 
Coll. Math. Societatis J\'anos Bolyai, {\bf 56} Differential Geometry, (1989), 231-252}

\bib{GM}{M. Godina, P. Matteucci,
{\it Reductive $G$-structures and Lie derivatives},
Journal of Geometry and Physics {\bf 47} (2003) 66--86
}

\bib{HolClass}{M. Berger, 
{\it Sur les groupes dÕholonomie des vari\'et\'es \`a connexion affine et des vari\'et\'es Riemanniennes},
 Bull. Soc. Math. France {\bf 83}, 279-330 (1955)}

\bib{HolClassExc}{S. Merkulov, L.Schwachh\"ofer
{\it ClassiÞcation of irreducible holonomies of torsion-free affine connections},
Annals of Mathematics, {\bf 150}(1) (1999), 77 - 149; [arXiv: math/9907206]}

\bib{Antonsen}{F.\ Antonsen, M.S.N.\ Flagga, 
{\it Spacetime Topology (I) - Chirality and the Third Stiefel-Whitney Class}, 
Int.\ J.\ Th.\ Phys.\ {\bf 41}(2), 2002}

\bib{KobaNu}{S.\ Kobayashi, K.\ Nomizu,
  {\it Foundations of differential geometry},
  John Wiley \& Sons, Inc., New York, 1963 USA}

\bib{Uni1}{L.Fatibene, M.Ferraris, M.Francaviglia, 
{\it New Cases of Universality Theorem for Gravitational Theories}, 
Classical and Quantum Gravity {\bf 27}, 165021 (2010); arXive: 1003.1617 
}

\bib{Uni2}{L.Fatibene, M.Ferraris, M.Francaviglia, 
{\it Extended Loop Quantum Gravity},
Classical and Quantum Gravity {\bf 27}, 185016 (2010); arXiv:1003.1619}

\bib{Alexandrov}{S.Alexandrov,
{\it On choice of connection in loop quantum gravity}, 
Phys.Rev. D65 (2002) 024011}

\bib{Livine}{S.Alexandrov, E.R.Livine,
{\it $SU(2)$ loop quantum gravity seen from covariant theory}
Phys.Rev. D67 (2003) 044009}

\bib{Holst}{S. Holst, 
{\it BarberoÕs Hamiltonian Derived from a Generalized Hilbert-Palatini Action}, 
Phys. Rev. D{\bf 53}, 5966, 1996}

\bib{semPI}{N. Bodendorfer, T. Thiemann, A. Thurn,
{\it New Variables for Classical and Quantum Gravity in all Dimensions III. Quantum Theory},
arXiv:1105.3705v1 [gr-qc]}

\bib{2times}{J. A. Nieto,
{\it Canonical Gravity in Two Time and Two Space Dimensions}, 
arXiv:1107.0718v3 [gr-qc]
}

\bib{NostroBI}{L. Fatibene,  M. Francaviglia, M. Ferraris,
{\it Inducing Barbero-Immirzi connections along SU(2) reductions of bundles on spacetime},
Phys. Rev. D, {\bf 84}(6),  064035 (2011)}

\bib{Maple}{K.Chu, C.Farel, G.Fee, R.McLenaghan, Fields Inst. Comm. {\bf 15}, (1997)}


\long\def\Old#1{}


\NormalStyle
\NormalStyle

\CollapseAllCNotes

\title{Do Barbero-Immirzi connections exist}
\moretitle{ in different dimensions and signatures?}

\author{L. Fatibene$^{1, 2}$, M. Francaviglia$^{1, 2, 3}$, S.Garruto$^1$}

\address{$^1$ Department of Mathematics, University of Torino (Italy)}

\moreaddress{$^2$ INFN- Sezione Torino; Iniz.~Spec.~Na12 (Italy)}

\moreaddress{$^3$ LCS - University of Calabria (Italy)}

\abstract
We shall show that no reductive splitting of the spin group exists in dimension $3\le m\le 20$ other than in dimension $m=4$.
In dimension $4$ there are reductive splittings in any signature. Euclidean and Lorentzian signatures are reviewed in particular and signature $(2,2)$ is investigated explicitly in detail.

Reductive splittings allow to define a global $\SU(2)$-connection over spacetime which encodes in an weird way the holonomy of the standard spin connection.
The standard Barbero-Immirzi (BI) connection used in LQG is then obtained by restriction to a spacelike slice.
This mechanism provides a good control on globality and covariance of BI connection showing that in dimension other than $4$ one needs to provide 
some other mechanism to define the analogous of BI connection and control its globality.
\endabstract

\NewSection{Introduction}

Barbero-Immirzi (BI) connection is used in LQG to describe gravitational field on a spacelike slice of spacetime; see \ref{Barbero}, \ref{Immirzi}. 
In standard literature it is obtained by a canonical transformation on the phase space of the spatial Hamiltonian system describing classical GR; see \ref{RovelliBook}.

The discussion about the possibility of defining a BI counterpart at the level of spacetime has been longly discussed  in literature (see  \ref{Samuel},  \ref{Thie2006}).
The discussion mainly focused on the possibility of obtaining the BI space connection by {\it restricting} a suitable BI spin connection defined globally over spacetime as a spacetime object.

We recently showed that the standard spatial BI connection can be in fact obtained by restriction on space of a spacetime $\SU(2)$-connection (see \ref{R1}) in spite of controversial opinions about such a possibility. 
Such a $\SU(2)$-connection is not though simply related to the spacetime spin connection; it is obtained by
a mechanism called {\it reduction} and its global properties can be controlled in view of an algebraic group-theoretical structure
called a {\it reductive group splitting} (see \ref{NostroBI}).

When one defines connections by {\it restriction} then constraints on the holonomy group of the restricted connection apply (see \ref{HolClass}, \ref{HolClassExc}) showing that standard spatial BI connection cannot be obtained directly by restriction from the spacetime spin connection.
However, such holonomic constraints disappear when the connection is defined by {\it reduction};
as a matter of fact any $\Spin(\eta)$-connection can be reduced to a  $\SU(2)$-connection.
Unfortunately, reduction produces an encoding of the holonomy of the original spin connection into the holonomy of the reduced connection; 
such an encoding is far from being trivial and it needs to be further investigated.

The standard BI connection defined in LQG exists because of a number of coincidences; first of all there exist
group embeddings $\io:\SU(2)\arr \Spin(4)$ and $\io:\SU(2)\arr \Spin(3,1)$ which are {\it reductive}.
Second, in dimension 4 a number of topological coincidences guarantee that any spin bundle over spacetime can be reduced to 
a $\SU(2)$-bundle under the mild hypotheses which are equivalent to the existence of global Lorentzian metrics and global spin structures (see \ref{KobaNu}, \ref{Antonsen}, \ref{NostroBI}). 
Finally, the dynamics can be written in terms of the BI connection by adding to the Hilbert action a term which is vanishing on-shell and not compromizing the classical sector; the modified action is called the {\it Holst} action (\ref{Holst}, \ref{R2}, \ref{Uni1})
and it provides a dynamically equivalent formulation of standard GR.

Of course, standard BI approach is not the only way to work out LQG. Different frameworks have been proposed (see \ref{Alexandrov} and \ref{Livine} just to mention some of them). Nor one can exclude other frameworks to control global properties of BI connection (see \ref{semPI}).
Still we have to stress that, to the best of our knowledge, the one based on {\it reductions} is the {\it only general framework known} (with the exception of some {\it ad hoc} method) to control global properties of standard BI connection at the full level of spacetime.

In this paper we shall consider possible extensions of BI construction by reduction to different signatures and dimensions.
We shall show that the construction basically works only in dimension $m=4$ in all signatures (at least for dimension $3\le m\le 20$).

In Section 2 we shall briefly review the reduction framework.
In Section 3 we shall briefly extend the framework to general dimensions.
In Section 4 we shall report some result about non-existence of reductive splitting with groups relevant 
in dimension $m$ for $m\le 20$.
In Section 5 we shall check directly reductive splittings in all signatures in dimension 4. 
The Euclidean and Lorentzian signature are well known. Relatively new is the case of Kleinian signature $\eta=(2,2)$. 
BI connection has been proposed and used in signature $(2,2)$ (see \ref{2times}); 
however, to the best of our knowledge the global properties of BI connections for signature (2,2) and its relation to a reductive splitting is new.

\NewSection{Reductive splittings}

In this Section we shall briefly consider the algebraic structure that enable us to reduce the connections.
Let us consider a principal bundle $P$ with group $G$ and a subgroup $i: H\arr G$.
Let us then assume and fix any {\it $H$-reduction} $(Q, \io)$  of $P$ given by 
$$
\begindc{\commdiag}[1]
\obj(110,80)[hQ3]{$Q$}
\obj(180,80)[hP3]{$P$}
\obj(110,30)[M2]{$M$}
\obj(180,30)[M3]{$M$}
\mor{hQ3}{M2}{}
\mor{hP3}{M3}{}
\mor{hQ3}{hP3}{$\io$}
\mor{M2}{M3}{}[\atleft, \solidline] \mor(110,33)(180,33){}[\atleft, \solidline]
\enddc
\fn$$
The existence of such a reduction usually imposes topological conditions on spacetime. 
In the standard situation of $G=\Spin(3,1)$ and $H=\SU(2)$ the bundle reduction is automatically ensured 
by standard physical requirements (essentially by existence of global spinors).



The group embedding  $i:H\arr G$ induces an algebra embedding $T_e i: \goth\arr \gotg$.
Let us define the vector space $V=\gotg/\goth$ so to have the short squence of vector spaces
$$
\begindc{\commdiag}[1]
\obj(10,20)[oL]{$0$}
\obj(50, 20)[h]{$\goth$}
\obj(100, 20)[g]{$\gotg$}
\obj(150, 20)[V]{$V$}
\obj(190, 20)[oR]{$0$}
\mor{oL}{h}{}
\mor{h}{g}{$T_e i$}
\mor{g}{V}{$p$}
\mor{V}{oR}{}
\cmor((150,15)(145,8)(125,6)(105,8)(100,15))
	\pup(125,1){$\Phi$}[\atleft, \dasharrow]
\enddc
\fn$$
where $\Phi: V\arr \gotg$ is a sequence splitting (i.e.~$p \circ \Phi= \id_V$) which always exists for sequences of vector spaces. 
Accordingly, one has  $\gotg\simeq \goth\oplus \Phi(V)$.

We say that $H$ is {\it reductive} in $G$ if there is an action $\la: H\times V \arr V$ such that $\ad(h)(\Phi(v))\equiv \Phi\circ \la(h, v)$
where $\ad(h): \gotg \arr \gotg$ is the restriction to the subgroup $H$ of the {\it adjoint action} of $G$ onto its algebra $\gotg$; see \ref{KobaNu}, \ref{Gatto}, \ref{GM}. 
In other words, the subspace $\Phi(V)\subset \gotg$ is invariant with respect to the adjoint action of $H\subset G$ on the algebra $\gotg$.

Let us stress that the vector subspace $\Phi(V)\subset \gotg$ is not required to be (and often it is not) a subalgebra; 
 one just needs the group embedding $i:H\arr G$.
A  bundle $H$-reduction $\io:Q\arr P$ with respect to a subgroup $H$ reductive in $G$ is enough to allow that 
{\it each} $G$-connection $\om$ on $P$ induces an $H$-connection on $Q$, which will be called the {\it reduced connection} (see \ref{R1} and \ref{NostroBI}).

\NewSection{Connections in Dimension $m>2$}

To fix notation let us consider  here spacetimes with dimension $m\equiv n+1>2$ and signature $\eta=(n,1)$;
the relevant spin groups are $\Spin(n)$ for space and $\Spin(n,1)$ for spacetime.
Accordingly, we are using signature $\diag(-1, 1, 1, \dots, 1)$ on $M$ so that the first coordinate $x^0$ corresponds to time.

Here both the groups are thought as embedded within their relevant Clifford algebra; see \ref{Book}.
The even Clifford algebras (where the groups' Lie algebras are embedded) are spanned by even products of Dirac matrices, here denoted by $\one, E_{\al\be}, E_{\al\be\ga\de}, \dots$ with $\al,\be,\dots= 0..n$.
The Clifford algebras are suitably embedded one into the other by 
$$
i_0: \calC(n)\arr \calC(n,1): E_{i_1\dots i_{2l}}\mapsto E_{i_1\dots i_{2l}}
\fn$$ 
with $i_1, i_2\dots= 1..n$. In other words, the lower dimensional Clifford algebra $\calC(n)$ is realized within the higher dimensional one $\calC(n,1)$ by means of even products of Dirac matrices, except $E_0$.
Such an algebra embedding restricts to a group embedding
$$
i:\Spin(n)\arr \Spin(n,1)
\fl{SpinEmbedding}$$

The corresponding covering maps allow to define  the embedding of $j:\SO(n)\arr \SO(n,1)$ which corresponds to rotations that fix the first axis, i.e. 
$$
j:\SO(n)\arr \SO(n,1): \la \mapsto \(\matrix{
1&0\cr
0& \la}\)
\fn$$

We have to show that the embedding \ShowLabel{SpinEmbedding} is reductive. Let us consider the sequence
$$
\begindc{\commdiag}[1]
\obj(10,20)[oL]{$0$}
\obj(70, 20)[h]{$\spin(n)$}
\obj(140, 20)[g]{$\spin(n,1)$}
\obj(210, 20)[V]{$V$}
\obj(270, 20)[oR]{$0$}
\mor{oL}{h}{}
\mor{h}{g}{$_{T_e i}$}
\mor{g}{V}{$_p$}
\mor{V}{oR}{}
\cmor((210,15)(205,8)(165,6)(145,8)(140,15))
	\pup(175,1){$_\Phi$}[\atleft, \dasharrow]
\enddc
\fn$$
The complement vector space $V$ is spanned by $E_{0i}$ and we fix
the splitting by setting
$$
\Phi: V\arr \spin(n,1): E_{0i}\mapsto E_{0i} + \frac[1/2] \be_i{}^{jk}E_{jk} 
\fn$$

One can write down the condition for which such a splitting is reductive, i.e.
$$
\la_i^l\be_l{}^{jk}= \be_i{}^{lm} \la^j_l \la^k_m
\fl{FiniteReductiveCondition}$$ 
which must hold true for any $\la\in \SO(n)$.
Then one can consider a $1$-parameter subgroup $\la(t)$ based at  the identity (i.e.~$\la(0)=\one$)
and the corresponding  Lie algebra element $\dot \la=\dot \la(0)$; 
the infinitesimal form of \ShowLabel{FiniteReductiveCondition} is then
$$
\dot \la_i^l\be_l{}^{jk}= \be_i{}^{lk} \dot \la^j_l +  \be_i{}^{jm}\dot \la^k_m
\fl{InfiniteReductiveCondition}$$
which must hold for any $\dot\la\in\so(n)\simeq \spin(n)$, i.e.~for any skew--symmetric matrix.

Then one should try to look for solutions of condition \ShowLabel{InfiniteReductiveCondition} that correspond to
reductive splittings, besides the trivial case $\be_i{}^{jk}=0$ which corresponds to no Immirzi parameter. Before searching for explicit solutions for $2\le n\le 19$ (i.e.~spacetime dimension $3\le m\le 20$)
let us consider few simple examples.

For $n=2$, Latin indices range in $i,j,\dots = 1,2$. The condition \ShowLabel{InfiniteReductiveCondition} specifies to
$$
\cases{
&\be_1{}^{12}=  \be_2{}^{12}\cr
&\be_2{}^{12}= - \be_1{}^{12}\cr
}
\fn$$
Hence one has $\be_1{}^{12}=  \be_2{}^{12}=0$, so that there is no reductive splitting other then $\be_i{}^{jk}=0$.

For $n=3$ (i.e.~$m=4$), Latin indices range in $i,j,\dots = 1,2,3$. The condition \ShowLabel{InfiniteReductiveCondition} 
has the only solution is $\be_i{}^{jk}= \be \ep_i{}^{jk}$ which spans reductive splittings (see see \ref{R1} and \ref{NostroBI}).
The constant parameter $\be$ is related to the standard Immirzi parameter.

One can immediately generalize that constructions in two classes of embeddings.
In both cases let us fix on $M$ signature $\eta=(r,s)$ (with $r+s=m$).
In the first case we take signature $\eta_{ab}=\retsme{s}{r}$  and consider the embedding
$$
i: \Spin(r,s-1)\arr \Spin(r,s) 
\fn$$
Accordingly, one is left with a  signature $\hat \eta=(r,s-1)=(k, l)$ on the {\it ``spatial'' leaf} of dimension $n=m-1= k+l$.
The  standard canonical form of signature $\hat \eta=(k,l)$ is fixed to be $\hat \eta_{ij}=\retsme{l=s-1}{k=r}$.

For notation convenience, in the second case we take signature $\eta_{ab}=\etsme{r}{s}$  and consider the embedding
$$
i: \Spin(r-1,s)\arr \Spin(r,s) 
\fn$$
Accordingly, one is left with a  signature $\hat \eta=(r-1,s)=(k,l)$ on the {\it ``spatial'' leaf} of dimension $n=m-1=k+l$.
The  standard canonical form of signature $\hat \eta=(k,l)$ is fixed to be $\hat \eta_{ij}=\etsme{k=r-1}{l=s}$.

In both cases we select the first axis as a fixed rotational axis and denote by $\eta_{ij}$ the standard canonical form of signature $\hat \eta=(k,l)$.

\NewSection{Non-existence of reductive splittings in dimension different from $m=4$}

In order to verify whether a reductive splitting occurs in an arbitrary dimension we must solve equations \ShowLabel{InfiniteReductiveCondition},
or better said the system obtained from \ShowLabel{InfiniteReductiveCondition} fixing and arbitrary $\dot \la\in \spin(n)$. 
Since the number of equations increases with the dimension of the space, it is difficult to find  solutions by direct calculations.
However, one can use Maple tensor package (see \ref{Maple}) to easily compute the solution of linear system \ShowLabel{InfiniteReductiveCondition} for any arbitrary (but fixed) dimension and signature.

First of all, one should look for the  general expression of the generators $\dot \la_i^l$ of the Lie algebra $\spin(k,l) \simeq \so(k,l)$.
Let us fix  the standard bilinear form $\hat \eta_{ij}=\etsme{k}{l}$ of signature $\hat \eta=(k,l)$; 
then the corresponding orthogonal group $\SO(\hat\eta)$ is the set of matrices defined by the relation:
$$ \la^i_k \hat\eta_{ij} \la^j_l = \hat\eta_{kl} 
\fn$$

The relation above can be read in the algebra as:
$$
\dot\la^i_k \hat\eta_{ij} + \hat\eta_{ki}\dot\la^i_j = 0 
\fl{AlgebraConditions}$$

It is easy to see that conditions \ShowLabel{AlgebraConditions} tell us that $\dot \la_i^l$ is a block matrix: 

$$\dot\la=\( \matrix
{A_{1} & B \cr
^{t} B& A_{2}
}\)
\fl{MatrixAlgebra}
$$
where $A_{1}$ and $A_{2}$ are skew-symmetric matrices, of dimension $k\times k$ and $l\times l$ respectively, while $B$ is an arbitrary $k\times l$ matrix. 
One can set generators of $\so(\hat\eta)$ to be matrices with all zero entries but two where $\pm 1$ is set according to 
\ShowLabel{MatrixAlgebra}.

Then equation \ShowLabel{InfiniteReductiveCondition} can be expanded along this basis of $\so(\eta)$ obtaining a system of 
$\frac[n^3/4](n-1)^2$ equations.
The unknowns $\be_k{}^{ij}$ are $\frac[n^2/2](n-1)$.
For any $n>2$ one has more equations than unknowns and has to compute the rank of the system to discuss solutions.
Of course computing the rank of the system obtained from \ShowLabel{InfiniteReductiveCondition} is rather difficult in general 
thus we shall analyze each case separately.

Of course, since the system is homogeneous, it cannot be inconsistent but it must have at least the trivial solution.
We aim to discuss whether, in some dimension, there are  solutions other than the trivial one. 

As we have seen above, in a fixed dimension $m$ and signature $\eta=(r,s)$ there are two ways of defining group embeddings, one fixing a time axis and one fixing a space axis.  So we have to check both of them.

We have obtained a computer-aided solution for  the system in all spacetime dimensions from $m = 3$ up to $m = 20$; in each dimension we considered any signature of spacetime
$\eta=(r,s)$ with $0\le r\le m$ and $s=m-r$; in each such dimension and signature we consider both cases, i.e.~fixing a time axis or a space axis.

[Of course, if $r=0$ one can only fix a time axis. Analogously, if $r=m$ (and $s=0$) one can only fix a space case.]

In all these cases (except for case $m = 4$ which will be analyzed in the next section)  none of the group splitting considered is reductive,  besides the trivial case $\be_k{}^{ij}=0$. 
Regardless the existence of bundle reductions, in these cases there is no canonical way of defining BI connections 
and one has to find out different mechanism (e.g.~resorting to embeddings involving different groups) to control global properties and covariance of BI connections (possibly changing the groups involved)
and to proceed to quantize {\it \'a la} loop.

\NewSection{Reductive splittings in dimension $m=4$}

Among the considered dimensions ($3\le m\le 20$), we found that only in $m = 4$ there are non-trivial reductive splittings. 
In dimension $m=4$ one has five signatures, three of them with 2 embeddings to be analyzed and two with one embedding only,
for a total of 8 embeddings to be considered.
In all these cases, it turns out to be that the splitting coefficients $\be_l{}^{jk}$ are proportional to the Levi-Civita symbol:
$$
\be_l{}^{jk} = \be \ep_{l}^{.jk}:=\be \hat \eta_{lm} \ep^{mjk}
\qquad\qquad
(\ga\in \R)
\fn$$
each using the relevant standard form $\hat \eta_{lm}$ according to the notation explained above.

Once $\be_l{}^{jk}$ are calculated we can directly verify from the definition that splittings in dimension four are all reductive.

First of all we shall define some useful notation:  let us set $\tau_{i} = \frac[1/2] \ep_{i}{}_\cdot^{j}{}_\cdot^{k} E_{jk}$ and $\si_{i} = E_{0i}$.
Since we shall have to compute products of $\tau_{i}$ it is convenient to write them in a closed form. One can verify that:
$$
\tau_{i}\tau_{j} = - \eta_{00} \eta \eta_{ij} \one - \ep_{ij}{}_\cdot^k \tau_{k}
\fn$$
where, by an abuse of language, we denote by $ \eta$ the determinant of $\eta_{ab}$.

Furthermore we can write the splitting $e_{k} = (-\al^3 E + \hat \be) \tau_{k}$, where $\hat\be= \be \hat\eta $ is a constant simply related to $\be$ and we set
$\al := \sqrt{\eta}$ (possibly imaginary) and $E: = \al E_{0123}$. 
Let us remark also that if $S\in \Spin(k,l)$, than it can be written as a linear combination of $\Spin(k,l)$ generators, namely
$$
S = a^{0} \one + a^{i} \tau_{i}
\fn$$
with inverse:
$$
S^{-1} = a^{0} \one - a^{i} \tau_{i}
\fn$$
under the constraint:
$$
(a_{0})^{2} + \eta_{00} \eta \left| \vec{a} \right|^{2} = 1
\fl{ConstraintSpin}
$$
which is the condition that defines spin group in $\calC^+(\eta)$.

With this notation we are ready to verify the splitting by applying directly the definition.
We have then to compute the adjoint action, restricted to $\Spin(k,l)$, on the bases $e_k$ of $\Phi(V)\subset \spin(r,s)$.
 One has:
$$\eqalign{S e_{k} S^{-1} & = (a^{0} \one + a^{i} \tau_{i}) (\al \eta E + \hat\be) \tau_{k} (a^{0} \one - a^{j} \tau_{j}) = \cr 
                                & = (\al \eta E +  \hat\be) (a^{0} \one + a^{i} \tau_{i}) (a^{0} \tau_{k} - a^{j} \tau_{k} \tau_{j}) = \cr
                                & = (\al \eta E +  \hat\be) (a^{0} \one + a^{i} \tau_{i}) (a^{0} \tau_{k} - a^{j} (-\eta_{00} \eta \hat\eta_{kj}\one
                                	- \ep_{kj}{}_\cdot^{l} \tau_{l})) = \cr
                                & = (\al \eta E +  \hat\be) ((a^{0})^{2}\tau_{k} + \eta_{00}\eta a_{k}^{.} a^{0}\one + a^{0}a^{j}
                                    \ep_{kj}{}_\cdot^{l}\tau_{l} + \cr
                                & \quad\quad + a^{0}a^{i} \tau_{i} \tau_{k} + a^{i}a^{.}_{k}\eta_{00}\eta \tau_{i} + 
                                	a^{i} a^{j} \ep_{kj}{}_\cdot^{l} \tau_{i} \tau_{l}) = \cr
			    & = (\al \eta E +  \hat\be) ((a^{0})^{2}\tau_{k} + \uline{\eta_{00}\eta a_{k}^{.} a^{0}\one} + \uuline{a^{0}a^{j}
                                    \ep_{kj}{}_\cdot^{l}\tau_{l}} + \uline{a^{0}a^{.}_{k} (-\eta_{00} \eta\one)} 
                                    -\uuline{a^{0}a^{i}\ep_{ik}{}_\cdot^{l} \tau_{l}} \cr
                                & \quad\quad + a^{i}a^{.}_{k}\eta_{00}\eta \tau_{i} + 
                                a^{i} a^{j} \ep_{kji} (-\eta_{00} \eta) 
                                     - a^{i} a^{j} \ep_{kj}{}_\cdot^{l} \ep_{il}{}_\cdot^{m} \tau_{m})=\cr
                                & = (\al \eta E +  \hat\be) \left((a^{0})^{2}\tau_{k} - 2a^{0}a^{j}
                                    \ep_{jk}{}_\cdot^{l}\tau_{l} + a^{m}a^{.}_{k}\eta_{00}\eta \tau_{m} 
                                     - a^{i} a^{j} \ep_{kj}{}_\cdot^{l} \ep_{il}{}_\cdot^{m} \tau_{m} \right)
}\fn$$

By using the contraction formula $\ep_{kjl} \ep_{i}^\cdot{}^{ml}  = \hat\eta_{ki} \hat\eta_{j}^{m} - \hat\eta^{m}_{k} \hat\eta_{ji}$  we can re-write $S e_{k} S^{-1}$ as:
$$
Se_{k}S^{-1} = l_{k}^{m} e_{m}
\fn$$
where
$$
l_{k}^{m} = \left((a^{0})^{2} - \eta \eta_{00} \vert \vec{a} \vert^{2} \right) \de_{k}^{m} + 2\eta \eta_{00} a^{m} a_{k}^{.}
- 2a^{0} a^{i} \ep_{ik}{}_{.}^{m}
\fl{OrthTransf}
$$

 If one uses \ShowLabel{ConstraintSpin} it is easy to see that \ShowLabel{OrthTransf} is an orthogonal transformation for $\hat \eta_{ab}$, namely, $\l^{m}_{i} \hat \eta_{mn} l^{m}_{j} = \hat \eta_{ij}$.
In this way we have been able to show that in dimension $m=4$ the splittings are reductive in all signatures.

\NewSection{Conclusions and perspectives}

We showed that for any dimension $3\le m= r+s \le 20$ all the embeddings
$$
i : \Spin(r-1, s)\arr \Spin(r,s)
\qquad
i : \Spin(r, s-1)\arr \Spin(r,s)
\fn$$
are {\it not} reductive except when $m=4$.

In $m=4$ they are all reductive for any choice of the signature, i.e.~for $0\le r\le m$.
In Euclidean signature the reductive splitting $i : \Spin(3)\arr \Spin(4)$ reproduces the standard BI connection used in the Euclidean sector.
In Lorentzian signature the reductive splitting $i : \Spin(3)\arr \Spin(3,1)$ reproduces the standard BI connection used in the Lorentzian sector.

The other signatures in dimension $m=4$ allow us to define a BI $\SU(2)$-connection on spacetime which produces  the
BI in Hamiltonian formalism by restriction. By this mechanism the global properties of the BI are under control and the holonomy encoding of the spin connection into the holonomy of the BI connection is manifest, though it surely deserves further investigations.

In dimension other than $4$ this mechanism cannot be used in order to guarantee the existence of global BI connections (or fields which behaves as connections under gauge transformations enforcing covariance of holonomic variables) and one  needs to rely on some other construction to quantize gravity as in LQG, possibly relying on some other group as suggested e.g. in \ref{semPI}.

\Acknowledgements
We wish to thank L.~Gatto, M.~Godina and P.~Matteucci for discussions about reductive splittings
and  A.~Smirnov for having pointed out to us the argument based on holonomy constraints.
We also thank C.~Rovelli for discussion about BI-connections.
We also acknowledge the contribution of INFN (Iniziativa Specifica NA12) the local research project 
{\it Metodi Geometrici in Fisica Matematica e Applicazioni} (2011) of Dipartimento di Matematica of University of Torino (Italy).
This paper is also supported by INdAM-GNFM.

\ShowBiblio

\end